\documentclass[10pt,journal]{IEEEtran}

\usepackage{graphics}
\usepackage{color,soul}
\usepackage{graphicx}
\usepackage{times}
\usepackage[numbers,sort&compress]{natbib}
\usepackage{hyperref}
\usepackage{subcaption}
\usepackage{fancyhdr}
\usepackage{booktabs}
\usepackage{multirow}
\usepackage{amssymb}

\newcommand{\takeaway}[1]{\textit{\textbf{Takeaway:}} \textit{#1}}

\usepackage{mdframed}
\definecolor{gray90}{gray}{0.90}
\definecolor{gray95}{gray}{0.95}
\newenvironment{summary}
{\vspace{-1em}\noindent\begin{mdframed}[backgroundcolor = gray90, linecolor = black, innerleftmargin=2mm, innerrightmargin=2mm]}
{\end{mdframed}}

\fancypagestyle{FirstPage}{
}

\begin{document}

\title{Evaluating Cross-Layer Interactions of QUIC, Encrypted DNS and HTTP/3: Design, Implementation and Performance}

\title{On Cross-Layer Interactions of QUIC, Encrypted DNS and HTTP/3: Design, Implementation and Evaluations}

\title{On Cross-Layer Interactions of QUIC, Encrypted DNS and HTTP/3: System Design and Implementation}

\title{On Cross-Layer Interactions of QUIC, Encrypted DNS and HTTP/3: System Design, Evaluations and Dataset}

\title{On Cross-Layer Interactions of QUIC, Encrypted DNS and HTTP/3: Design, Evaluation and Dataset}

\author{\IEEEauthorblockN{Jayasree Sengupta\IEEEauthorrefmark{1},
Mike Kosek\IEEEauthorrefmark{2},
Justus Fries\IEEEauthorrefmark{2}, 
Simone Ferlin\IEEEauthorrefmark{3}, and
Vaibhav Bajpai\IEEEauthorrefmark{4}}\\
\IEEEauthorblockA{\IEEEauthorrefmark{1}CISPA Helmholtz Center for Information Security, Germany}
\IEEEauthorblockA{\texttt{[jayasree.sengupta@cispa.de]}}\\
\IEEEauthorblockA{\IEEEauthorrefmark{2}Technical University of Munich, Germany}
\IEEEauthorblockA{\texttt{[kosek@in.tum.de $|$ justus.fries@tum.de]}}\\
\IEEEauthorblockA{\IEEEauthorrefmark{3}Red Hat and Karlstad University, Sweden}
\IEEEauthorblockA{\texttt{[simone@ferlin.io]}}\\
\IEEEauthorblockA{\IEEEauthorrefmark{4}Hasso Plattner Institute, Germany}
\IEEEauthorblockA{\texttt{[vaibhav.bajpai@hpi.de]}}
}

\maketitle
\IEEEpeerreviewmaketitle

\begin{abstract}
Every Web session involves a DNS resolution. While, in the last decade, we witnessed a promising trend towards an encrypted Web in general, DNS encryption has only recently gained traction with the standardisation of DNS over TLS (DoT) and DNS over HTTPS (DoH). Meanwhile, the rapid rise of QUIC deployment has now opened up an exciting opportunity to utilise the same protocol to not only encrypt Web communications, but also DNS. In this paper, we evaluate this benefit of using QUIC to coalesce name resolution via DNS over QUIC (DoQ), and Web content delivery via HTTP/3 (H3) with 0-RTT. We compare this scenario using several possible combinations where H3 is used in conjunction with DoH and DoQ, as well as the unencrypted DNS over UDP (DoUDP). We observe, that when using H3 1-RTT, page load times with DoH can get inflated by $>$30\% over fixed-line and by $>$50\% over mobile when compared to unencrypted DNS with DoUDP. However, this cost of encryption can be drastically reduced when encrypted connections are coalesced (DoQ + H3 0-RTT), thereby reducing the page load times by 1/3 over fixed-line and 1/2 over mobile, overall making connection coalescing with QUIC the best option for encrypted communication on the Internet.

\end{abstract}


\section{Introduction}
\thispagestyle{FirstPage}

Over the last decade, with the increased privacy awareness 
amongst individuals, the Web slowly started becoming encrypted~\cite{tls.adoption,meliaconext2018}.
However, encrypted DNS has only recently gained traction with the standardisation of DNS over TLS (DoT)~\cite{rfc7858} and DNS over HTTPS (DoH)~\cite{rfc8484}, where in today's Internet unencrypted DNS resolution using DNS over UDP (DoUDP) remains the default~\cite{deccio2019dns}.
Hence, despite the encryption of the actual Web content, the browsing behaviors of individuals can still be observed, enabling third parties to create trackable user profiles~\cite{KimZ15,KirchlerHLK16,LiMGLZLG18,connection.oriented.dns}.

To counter this problem, today's browsers offer to encrypt DNS traffic using DoH~\cite{dns.privacy.resolvers}, enabling users to opt-in into encrypted DNS with a public DNS resolver~\cite{doan2021resolvers} of their choice.
While DoH adds privacy to the DNS, hence enabling \textit{Web Privacy By Design}, it remains rarely used, and is inherently limited by the underlying protocols:
Multiple studies evaluate the impact of DoH and DoT on Web performance, finding that they are constrained by head-of-line blocking of the TCP connection, as well as the multiple round-trips required for the handshake of the TCP and TLS sessions~\cite{doan2021measuring,encrypted.dns.be.fast,lu2019dns,doh.round.the.world,hounsel2020dns,cuadrado2019dns,borgolte2019dns}.

\begin{figure}[!t]
    \centering
    \begin{subfigure}[t]{0.5\textwidth}
        \centering
        \includegraphics[width=0.9\columnwidth]{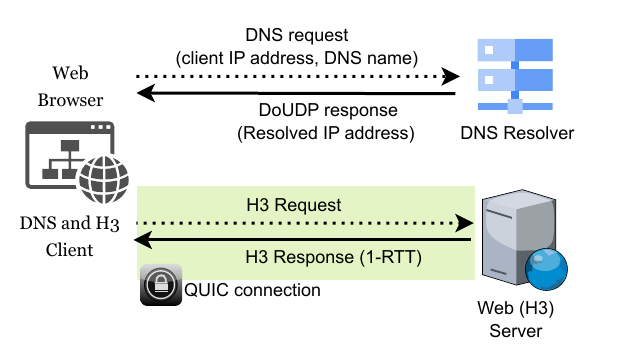}
        \caption{Existing mechanism of Web Browsing with QUIC whereby a DNS request gets resolved over un-encrypted DoUDP followed by an encrypted HTTP/3 session.}
        \label{fig:Image13a}
    \end{subfigure}%
    \vspace{1em}
    \begin{subfigure}[t]{0.5\textwidth}
        \centering
        \includegraphics[width=0.9\columnwidth]{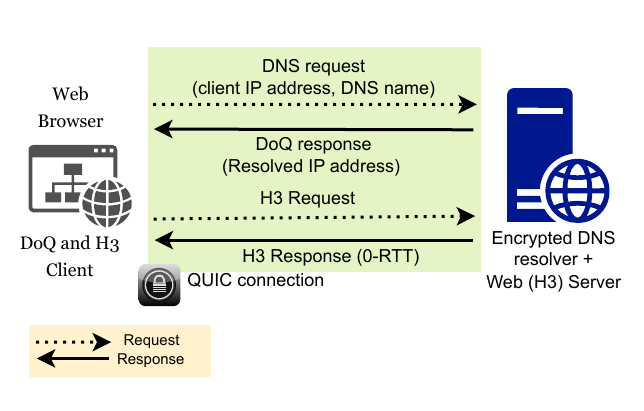}
        \caption{Proposed mechanism of Web Browsing whereby QUIC is used to coalesce name resolution with DoQ and Web content delivery with H3 $0-$RTT over a single QUIC connection.}
        \label{fig:Image13b}
    \end{subfigure}
    \caption{\small \sl Web Browsing over different unencrypted and encrypted DNS protocols using both H3 $0$-RTT and H3 $1$-RTT combinations.}
    \vspace{-1.5em}
\end{figure}

To overcome these inherent limitations of TCP and TLS, the QUIC transport protocol has recently been standardized, offering multiplexing support to address head-of-line blocking, and overcoming the handshake limitations by combining the transport and encryption handshake into a single round-trip~\cite{rfc9000}.
Moreover, QUIC can also leverage $0-$RTT in order to send application data within the first round-trip, effectively nullifying the handshake overhead altogether.
QUIC was designed in tandem with HTTP/3 with focus on the encrypted Web:
While H3 leverages QUIC as a transport protocol, requests can be multiplexed over a single QUIC connection, greatly reducing the overhead of HTTP/2 and HTTP/1.1 which are required to establish multiple TCP and TLS sessions in order to avoid head-of-line blocking \cite{Kosek_COMMAG'21}.
Hence, recent studies show that H3 improves over HTTP/2, finding reduced page load times (PLTs) for H3 while being less affected by packet loss and delay~\cite{quicPep,tanya2022}, yet highlighting the importance of configuration choice for the performance of QUIC~\cite{yu2021}. Moreover, encrypted DNS also benefits from QUIC, where the recently standardized DNS over QUIC (DoQ)~\cite{rfc9250} improves over DoH and DoT~\cite{doqWebPerf}\cite{doqpam2022}. Evaluating the impact on Web performance, it is shown that DoQ improves over DoH with up to 10\% faster page loads on simple Web pages, and DoQ resulting in only 2\% slower page loads in comparison to DoUDP on complex web pages. 

Hence, QUIC greatly improves on the notion of \textit{Web Privacy By Design}: where DoQ primarily benefits from faster handshakes, H3 avoids multiple handshakes by multiplexing requests over a single connection. Both protocols improve within their own layers, but the combination of DoQ and H3 significantly improves over DoH with HTTP/2. A typical Web browsing scenario over these protocols is depicted in Fig. \ref{fig:Image13a}.

However, even when using QUIC for both DoQ and H3, the improvements are still uncoupled. Yet, CDN providers like Cloudflare offer both public DNS services using DoQ and Web content delivery using H3 on the same edge infrastructure~\cite{cf_load_balancer}:
Consequently, DNS resolution using DoQ, and preceding H3 requests to a web page hosted by the same CDN, will both be served using QUIC from the same infrastructure, offering optimization potential. The fresh H3 request to the web server happens over the same QUIC connection. This is exactly where our proposed QUIC connection coalescing is applicable as shown in Fig. \ref{fig:Image13b}. For example, Cloudflare can majorly benefit from their existing setup to utilise QUIC to coalesce name resolution via DoQ and simultaneously execute Web content delivery using H3 with $0$-RTT. By doing so, the Web communication is not only private but also becomes faster by reusing the same underlying QUIC connection. Overall, we provide two main contributions:

\begin{itemize}
        \item[$\blacksquare$] 
        \noindent\textbf{Measurement Method --} We evaluate the cross-layer interactions of QUIC, DNS, and H3, analyzing the benefits of using QUIC to coalesce name resolution with DoQ and Web content delivery with H3 $0-$RTT. We hereby present a measurement setup (see: §~\ref{sec:methodology}) that automates DNS resolution and Web browsing while emulating network conditions of a user at the edge based on real-world datasets for both fixed- and mobile-access network technologies.     
    \item[$\blacksquare$]     
    \noindent\textbf{Findings --} We show (see: §~\ref{sec:evaluation}) that page load times using DoH can get inflated by $>$30\% over fixed-line and by $>$50\% over mobile when compared to unencrypted DNS with DoUDP, reflecting the cost of encrypted DNS using DoH~\cite{Naylor_CoNext'14}. Taking \textit{Web Privacy By Design} to the next level, we coalesce DoQ and H3 $0-$RTT connections, thereby reducing page load times by 1/3 over fixed-line and 1/2 over mobile in comparison to existing setup, overall making connection coalescing with QUIC the best option for encrypted communication on the Internet. In order to enable the reproduction of our findings, we have made the raw data of our measurements as well as the analysis scripts and supplementary files available\footnote{\url{https://github.com/justus237/DoQ-H3-analysis}}. 
        
\end{itemize}

This paper builds on our earlier work \cite{WebPrivacy_IFIP'23}. In this paper, we have added substantial background material, including a review (see: §~\ref{sec:background}) of recent web performance testing, monitoring and related methods over all three encrypted DNS protocols (DoT, DoH and DoQ), QUIC and HTTP/3. We have added further details to our methodology with illustrations of the measurement setup (see: §~\ref{sec:methodology}) to aid the readership. In addition, we have added new results, encompassing a detailed analysis (see: §~\ref{sec:website_categories}) on two categories of websites: (a) HTML Page with Javascript (b) HTML Page with Javascript and CSS. We have also added a new section further highlighting the broader implications (see: §~\ref{sec:discussion}) of our work, and discussing new research directions. Our results establish that QUIC connection coalescing is the best option for encrypted communication on the Internet, however, performance gains vary depending on the website and access technology combination used. Towards the end, we discuss §~\ref{sec:limitations} limitations and future scope, followed by the concluding remarks in §~\ref{sec:conclusion}.


\section{Background and Related Work \label{sec:background}}
In the following we introduce the three main protocols studied in this paper: DNS with its three most recent secured variations - DoT, DoH and DoQ - HTTP3, and QUIC.


\subsection{DNS protocol}

In essence, the DNS protocol is the Internet's phonebook, where a user asks a DNS resolver to translate a human-readable domain name, e.g., business.com into a machine-readable IP address and vice-versa. The DNS protocol typically uses port 53, also known as Do53, and supports unencrypted queries over both UDP and TCP protocols. Under the hood, the DNS protocol is a distributed system composed of a global network of nameservers organised as an hierarchical database of resource records. Typically, a user sends out a request for resource record to a recursive resolver, typically operated by network providers, acting as a proxy. If the resource record is already in this resolver's cache, the reply is sent straight back to the user. If it is not, the resolver will, starting from the root, traverse the hierarchical tree of nameservers until it receives an authoritative answer to the user's resource record request. 

In 1983, when the DNS protocol was introduced, privacy did not have the same consideration it has today~\cite{rfc1034, rfc1035}. For this reason, DNS messages are typically sent in plain text to the recursive resolvers. Unencrypted DNS messages reveal a great deal of the user's behaviour in the Internet, allowing anyone on the path between the user and nameserver to eavesdrop or make use of DNS for distributed Denial-of-Service (DDoS) attacks. In today's Internet, the DNS protocol also became a critical piece, due to its relevance for Content Distribution Networks (CDNs) for traffic redirection. With the development of Internet censorship and surveillance mechanisms, privacy considerations has been inevitably included in modern protocol standards' development, which includes more recent attempts to secure the DNS protocol, namely, DNS over TLS (DoT), DNS over HTTPS (DoH), and DNS-over-QUIC (DoQ).


DoT~\cite{rfc7858} establishes a TLS session between the user and the recursive resolver on port 853, to exchange subsequent encrypted DNS queries and responses. This portion of the DNS request, i.e., the path between user and the recursive resolver, can easily be associated with individual users. For this reason, but not limited to this portion of the DNS request, DoT can be seen as an attempt to primarily provide privacy for the portion of the DNS request between the user and the recursive resolver. DoH~\cite{rfc8484} runs atop TCP on port 443, which is the standard port for HTTPS, i.e., Internet traffic. By using HTTPS, DoH traffic inherently looks like any other encrypted Internet traffic. Thus, DoH has been considered more robust against censorship mechanisms or port-based firewalls compared to DoT. Then, DoH sends DNS requests in an HTTP GET request on HTTPS default port 443. DoQ~\cite{rfc9250} is a third attempt to improve DNS request privacy and minimize latency by leveraging QUIC as the underlying protocol. Although encrypted DNS protocols such as DoT and DoH are already deployed and in use, they suffer from shortcomings due to being based on TCP. In other words, although DoQ carries privacy properties similar to DoT and DoH, the latency characteristics of DoQ is more similar to the unencrypted DoUDP.

With several encrypted DNS standards available, research has been looking at unencrypted DNS~\cite{truncation, dotcp} and also comparing it with DoT, DoH and, more recently, DoQ (see: Table \ref{tab:literature}). In 2021,~\cite{garcia2021large} shows that although the amount of Internet traffic for encrypted DNS was not growing, there was a growing number of DoH servers available - for benign and malicious purposes.~\cite{kambourakis.2022.encrypted} provides an overview of DNS encryption proposals, discussing the value of the protection dependent on the trust of end users in the DNS resolvers. 
In~\cite{1041066}, the authors perform a trace file analysis with DNS traffic over two research institute networks looking at the performance of DNS requests, failures, errors, caching effectiveness.~\cite{encrypted.dns.be.fast} studies the performance of encrypted DNS versus unencrypted DNS in home networks, where DoT obtained lower latency compared to DNS whereas DoH had significant performance variation depending on the recursive resolver. Then,~\cite{doh.round.the.world} confirmed that performance of DoH varies, looking at geographic differences compared to unencrypted DNS. In~\cite{doqpam2022}, the authors looks at DoQ, showing a steady increase in adoption, with a good portion of the measurements indicating higher handshake times, however, with DoQ still outperforming DoT and DoH. Further,~\cite{lyu.2023.survey.dns} surveys the DNS encryption standards and literature between 2016 to 2021 looking at their adoption status, performance, benefits, and security issues. Then, the authors show the current landscape, how encrypted DNS is misused by malware, and also highlight DNS traffic inference techniques currently available. In~\cite{houser.2019.dot.leakage} the authors also look at how much traffic analysis can deduct from DoT messages, confirming that information leakage is possible even when DoT messages are padded.

\subsection{QUIC protocol}
The most recent transport layer revolution has been undoubtedly the QUIC protocol~\cite{rfc9000}. With the promise of being more simply extendable, maintainable and deployable, QUIC is a connection-oriented, end-to-end encrypted transport protocol based on UDP. 

With growing interest for QUIC in general, there has been research (see: Table \ref{tab:literature}) evaluating the protocol considering different network scenarios and technologies such as wireless, satellite networks and IoT, server and client stacks, configuration and location. Already in 2017,~\cite{quicWeb2017} compares QUIC with respect to the network, the website structure and involved end-to-end actors. Then,~\cite{rueth.poese.dietzel.hohlfeld.2018} confirmed that QUIC traffic already in 2018 accounted to up to 9\% of the Internet traffic.

In~\cite{fernandez.2020.mqtt}, the authors study the performance of the Message Queuing Telemetry Transport Protocol (MQTT) over QUIC, where the authors confirm good performances for typical IoT use cases.~\cite{choudhary.2020.quic} proposes two different cross-layer approaches to compares against QUIC over wireless networks, while~\cite{manzoor2020performance} looks at QUIC performance in wireless mesh networks. In~\cite{yu2021}, the authors compare the performance of QUIC and TCP against production servers hosted by Google, Facebook, and Cloudflare under several network conditions, applications, and client implementations, reporting performance benefits of QUIC largely linked to the QUIC server and client configurations such as congestion control and stack tuning.

Then,~\cite{tanya2022} evaluates QUIC performance over Internet transfers, cloud storage, and video applications, and it compares it against TLS/TCP. The authors confirm lower latencies for Internet transfers over QUIC, in cloud storage with certain file sizes, and with video streaming. In~\cite{nawrocki.2022.tls.quic}, the authors look at QUIC connection setup performance, more specifically at the the size and compression of TLS certificates, due to the impact in the handshake phase. Finally,~\cite{endres.2022.quic.performance} evaluates the performance of several QUIC implementations over several emulated and real-world geostationary satellite links. the authors report poor performance for QUIC, specially when there is packet loss.

\subsection{HTTP3 protocol}
The Hypertext Transfer Protocol (HTTP) is used to access the vast majority of services on today's Internet. The protocol was born in the early 90s with the goal to allow multimedia content and hyper-textual document transfers over the Internet. HTTP/1.1 standardized version came out in 1997~\cite{rfc2616}. In HTTP/1.1, only one resources can be in-flight on the underlying TCP connection, holding up all further resources behind it until it is fully downloaded. This is more generally known as Head-Of-Line (HOL) blocking. As the resources avalable in the Internet grew in size over the years, to achieve better page loading performance, Internet browsers started opening up several, up to six, parallel HTTP/1.1 connections per domain.

\begin{table}[!t]
\centering
\caption{\small \sl Existing research on Encrypted DNS, QUIC and HTTP/3}
\label{tab:literature}
\resizebox{0.9\columnwidth}{!}{%
\begin{tabular}{@{}clr@{}}
\toprule
\textbf{Protocols} & \textbf{Research Focus} & \textbf{References} \\ \midrule
\multirow{4}{*}{\begin{tabular}[c]{@{}c@{}}Encrypted\\ DNS\end{tabular}} & \begin{tabular}[c]{@{}l@{}}Performance comparison\\ of DNS protocols\end{tabular} & \cite{truncation, dotcp,1041066} \\ \cmidrule(l){2-3} 
 & \begin{tabular}[c]{@{}l@{}}Measurement on DNS\\ adoption\end{tabular} & \cite{garcia2021large,doqpam2022} \\ \cmidrule(l){2-3} 
 & \begin{tabular}[c]{@{}l@{}}DNS encryption and its\\ performance\end{tabular} & \cite{kambourakis.2022.encrypted,encrypted.dns.be.fast,doh.round.the.world,lyu.2023.survey.dns} \\ \cmidrule(l){2-3} 
 & \begin{tabular}[c]{@{}l@{}}Security analysis of\\ DNS protocols\end{tabular} &  \cite{lyu.2023.survey.dns,houser.2019.dot.leakage}\\ \midrule
\multirow{6}{*}{QUIC} & Deployment and adoption & \cite{quicWeb2017,rueth.poese.dietzel.hohlfeld.2018} \\ \cmidrule(l){2-3} 
 & QUIC with IoT & \cite{fernandez.2020.mqtt} \\ \cmidrule(l){2-3} 
 & \begin{tabular}[c]{@{}l@{}}Performance of QUIC\\ over different networks\end{tabular} & \cite{choudhary.2020.quic,manzoor2020performance,yu2021} \\ \cmidrule(l){2-3} 
 & \begin{tabular}[c]{@{}l@{}}QUIC's performance over\\ different workloads\end{tabular} & \cite{tanya2022} \\ \cmidrule(l){2-3} 
 & QUIC and TLS interplay & \cite{nawrocki.2022.tls.quic} \\ \cmidrule(l){2-3} 
 & QUIC over satellites & \cite{endres.2022.quic.performance} \\ \midrule
\multirow{6}{*}{HTTP/3 (H3)} & Resource Multiplexing & \cite{marx.2020.http2.vs.http3} \\ \cmidrule(l){2-3} 
 & \begin{tabular}[c]{@{}l@{}}HTTP Adaptive \\ Streaming (HAS) over H3\end{tabular} & \cite{lorenzi.2021.http3.abr} \\ \cmidrule(l){2-3} 
 & H3 with Lighthouse & \cite{saif2021} \\ \cmidrule(l){2-3} 
 & \begin{tabular}[c]{@{}l@{}}H3 adoption and\\ performance measurement\end{tabular} & \cite{trevisan.2021.http3.performance,perna.2022.http3.performance,gupta.2022.http3.realworld} \\ \cmidrule(l){2-3} 
 & H3 over LEO satellites & \multicolumn{1}{l}{} \cite{quicPep}\\ \cmidrule(l){2-3} 
 & H3 with IoT & \multicolumn{1}{l}{} \cite{saif.2021.http3.mqtt}\\ \bottomrule
\end{tabular}%
}
\end{table}

In 2014, HTTP's next version (HTTP/2)~\cite{rfc7540} was proposed with substantial changes in how data is framed and transported. As such, one of HTTP/2 main goals was to implement multiplexing of resources over a single underlying TCP connection. To achieve this goal, the protocol divides resource payloads into smaller uniquely-identified prefixed chunks, thus, allowing multiple resources on the wire.

Since 2022, HTTP/3~\cite{rfc9114} is the most recent version of HTTP and it promises performance and security improvements compared to HTTP/2. While HTTP/3 semantics and high-level features of HTTP/2 are kept intact, some core protocol aspects have been substantially reworked~\cite{marxquicly}. Beyond the replacement of the underlying transport protocol from TCP to QUIC, HTTP/3 comes with more efficient header compression, and advanced security features based on TLS 1.3. 

Meanwhile, research (see: Table \ref{tab:literature}) quantifying the benefits of HTTP3 in terms of Quality of Experience (QoE), HTTP features, different applications, and different network scenarios such as IoT, mobile and satellite networks has emerged with different outcomes: ~\cite{marx.2020.http2.vs.http3} compares resource multiplexing prioritization between HTTP2 and HTTP3 protocols.~\cite{lorenzi.2021.http3.abr} investigates HTTP Adaptive Streaming (HAS) over HTTP3, proposing an optimization to the Adaptive Bitrate (ABR) algorithm using HTTP3 request cancellation, and~\cite{saif2021} looks at diverse HTTP3 metrics with Lighthouse. 

In~\cite{trevisan.2021.http3.performance} the authors run a measurement study looking at HTTP3 adoption and performance, where at the time it testified its benefits limited to a few scenarios with high latency or poor bandwidth. Later,~\cite{perna.2022.http3.performance} revisited the topic with a slight different outcome: While confirming that the benefits of HTTP3 were more visible in high latency scenarios and also mobile networks, the did not observe any improvements with video streaming. In~\cite{gupta.2022.http3.realworld}, authors look at Quality of Experience (QoE) and the impact of local connectivity, server location and server software between HTTP2 and HTTP3, where they confirm better performance of HTTP3 over HTTP2 in challenging networking conditions.

Then,~\cite{quicPep} looks at HTTP3 performance in Low-Earth Orbit (LEO) satellites with and without Performance Enhancing Proxies (PEP), where the authors indicate better HTTP3 performance with and without proxy compared to its predecessors. Finally,~\cite{saif.2021.http3.mqtt} looks at the MQTT IoT protocol over HTTP3 indicating that they could save one RTT to publish messages to the broker, which in typical high-latency or low-power IoT environments, is significant.

\section{Methodology}
\label{sec:methodology}

\begin{figure}[!t]
    \centering
    \includegraphics[width=0.9\linewidth]{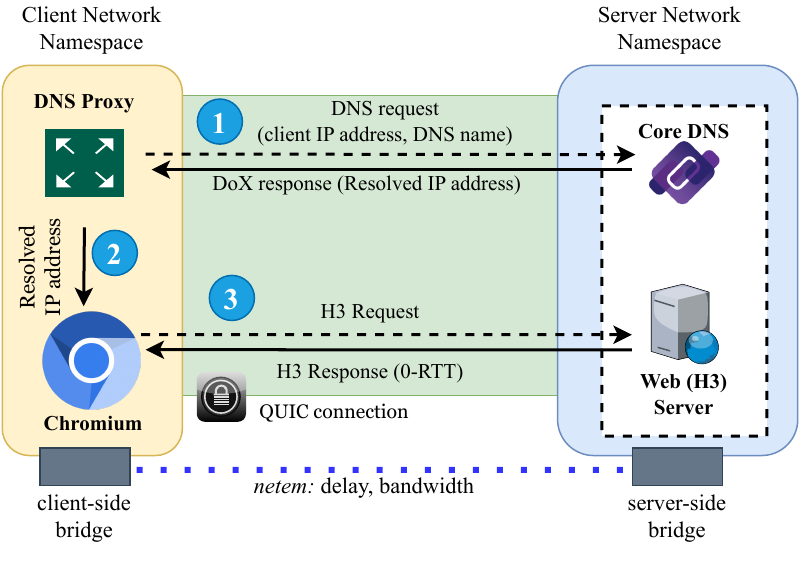}
    \caption{\small \sl Measurement setup used to evaluate QUIC connection coalescing using DoQ + H3 $0-$RTT. The setup automates DNS resolution and Web browsing while emulating network conditions, such as delay and bandwidth    of a user at the edge.}
    \label{fig:methodology}
    \vspace{-1em}
\end{figure}

To evaluate QUIC connection coalescing using DoQ + H3 $0-$RTT, our measurement setup (see: Fig. \ref{fig:methodology}) automates DNS resolution and Web browsing while emulating network conditions of a user at the edge. It is based on real-world datasets for both fixed and mobile-access network technologies.
Moreover, we compare this optimized approach to different combinations of H3 in conjunction with DoH and the unencrypted DoUDP due to their prevalence in today's browsers. To this end, the measurement setup decouples the DNS resolution from the actual web page loading on the client side, where the DNS and the H3 server run in the same process on the server side; as a design choice, we measure one DNS resolution to normalise the impact of DNS across different websites (see §\ref{sec:limitations}).

The measurement scenario is web browsing where \textit{Chromium}~\cite{chromium} is used to measure page load times of three categories of web pages: an HTML page (\href{http://www.example.org/}{example.org}), an HTML page with javascript assets (\href{https://www.wikipedia.org/}{wikipedia.org}) and an HTML page with javascript assets, CSS and cookies (\href{https://www.instagram.com/}{instagram.com}).
These web pages were downloaded on June, 2022 and are specifically chosen since they require only a single domain resolution to fully fetch the web page, i.e., all resources are fetched from the same host, and all HTTP requests are sent to it.
The websites are cloned and provided by \textit{quic-go} H3 server with gzip compression for all data including html.
To access a web page, first the domain name of the web page requested is resolved using DoQ, DoH, or DoUDP.
Following, H3 is used to connect to the resolved IP address in order to directly fetch the content and render it within the browser. 
During this step, QUIC connection coalescing is simulated by using a QUIC $0-$RTT handshake within \textit{Chromium’s} H3 request, i.e., sending the HTTP request in conjunction with the first QUIC handshake packet. 

The setup is encapsulated in Linux network namespaces, enabling separating client and server into different network domains. Following this, different network conditions are simulated using \texttt{netem} for fiber, cable, DSL and 4G.
For 4G, two variations are used: 4G with good signal quality (referred to as 4G), as well as 4G with medium signal quality (4G medium).
Table \ref{tab:Table1} shows the delay as well as bandwidth values that are applied for the different scenarios which are based on empirical data:
FCC’s Measuring Broadband America dataset \cite{mba2022} is used to represent the fixed broadband scenarios, whereas the ERRANT dataset \cite{TKG_CN'20,Midoglu_Infocom'18} is used for mobile wireless access technologies.
The delays and bandwidth are controlled using \texttt{netem}, where delay is always meant in the sense of two-way delay, i.e., the round-trip time (RTT), where the on-way delay is assumed to be symmetrical.

To enable this setup, we used \textit{Chromium 102.0.4991.0} and \textit{CoreDNS 1.7.0}~\cite{adguardcoredns}, where several changes were made to both these open source tools.
\textit{CoreDNS} was extended to additionally run an H3 server in order to share TLS information, resulting in an executable that runs both servers with the same certificates and \textit{session ticket} keys. 
Moreover, \textit{Chromium} was modified to support importing and exporting TLS session information, enabling $0$-RTT and TLS session resumption following browser restarts. The used machine was an Ubuntu 18.04 with Kernel 5.4.0, featuring 2 Intel Xeon E5-2643 6-Core CPUs and 128GB of RAM. It is to be noted that during the evaluation, we did not find any limitation by the hardware.

\section{Evaluation}
\label{sec:evaluation}

In order to evaluate QUIC connection coalescing, we first investigate the interaction of QUIC with DoQ and H3 in §~\ref{sec:3.1}, followed by an evaluation of the overhead of DoQ and DoH in comparison to the unencrypted DoUDP in §~\ref{sec:3.2}.
Finally, we perform a detailed analysis of the web performance for the combination of all three DNS protocols with H3 $1-$RTT as well as $0-$RTT, highlighting the benefits of QUIC connection coalescing in §~\ref{sec:3.3}. Our goal is to observe how different access technologies influence the behavior of these protocols, but not to evaluate a representative mix of internet access connections. In our dataset, all these protocol combinations have a sample size of 57,436. The different access technology scenarios are not distributed evenly due to measurement interruptions. The sample sizes are as follows: fiber 68,934, DSL 68,928, 4G 68,922, cable 68,916 and 4G medium 68,916. For the same reason, the sample sizes for the measurements are also not distributed evenly: \href{http://www.example.org/}{example.org} 114,924, \href{https://www.wikipedia.org/}{wikipedia.org} 114,882 and \href{https://www.instagram.com/}{instagram.com} 114,810.


\begin{table}[!t]
\centering
\caption{\small \sl Average values obtained from FCC’s Measuring Broadband America and ERRANT datasets}
\label{tab:Table1}
\resizebox{0.7\columnwidth}{!}{%
\begin{tabular}{@{}lrrr@{}}
\toprule
\textbf{\begin{tabular}[c]{@{}c@{}}Access \\ Technology\end{tabular}} & \textbf{\begin{tabular}[c]{@{}c@{}}Delay\\ (ms)\end{tabular}} & \textbf{\begin{tabular}[c]{@{}c@{}}Download\\ (Mbps)\end{tabular}} & \textbf{\begin{tabular}[c]{@{}c@{}}Upload\\ (Mbps)\end{tabular}} \\ \midrule
Fibre & 14.8 & 99.9 & 109.1 \\
Cable & 25.2 & 165.1 & 11.6 \\
DSL & 42.4 & 10.7 & 0.8 \\
4G & 91.9 & 54.0 & 21.2 \\
4G medium & 104.5 & 28.7 & 4.2 \\ \bottomrule
\end{tabular}%
}
\end{table}

\subsection{On QUIC's Interaction with Application Layer Protocols}
\label{sec:3.1}

\begin{figure}[!t]
    \centering
    \includegraphics[width=1\linewidth]{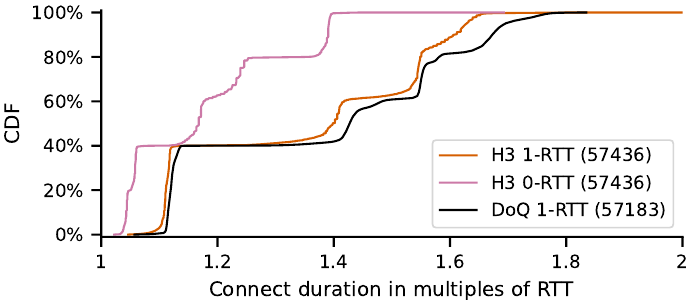}
    \caption{\small \sl CDF of the QUIC handshake connect duration H3 for 1-RTT and 0-RTT, as well as DoQ 1-RTT for all scenarios. The values are normalized by the delay that was applied during the measurement to show how these metrics scale with round-trips.}
    \label{fig:Image11}
    \vspace{-1em}
\end{figure}

Within this section we illustrate how the QUIC handshake interacts with H3 as well as its scaling capability over various network conditions. 
As part of the evaluation, Fig. \ref{fig:Image11} shows two relevant metrics for H3: \textit{connect} duration (i.e., \textit{connectEnd} - \textit{connectStart}) 
 and DoQ QUIC handshake duration measured in the DNS proxy. Here, \textit{‘connectStart’} signifies the timestamp immediately before the user starts establishing the connection to the server in order to retrieve the resource. In this experiment, the user establishes TCP and TLS sessions. On the other hand, \textit{‘connectEnd’} defines the timestamp immediately after the browser finishes establishing the connection to the server for retrieving the resource. The \textit{connect} duration is measured for both H3 with a 0-RTT and 1-RTT QUIC handshake. It is observed that H3 1-RTT \textit{connect} times appear to roughly correspond to DoQ handshake times. This was verified by looking at \textit{netlogs} and calculating the timespan between the client sending the initial and the last handshake packet (i.e., the \texttt{FIN} message), which appears to be at most around one millisecond lower than the reported \textit{connect} time. The \texttt{FIN bit (0x01)} of the frame type is set on frames that contain the final offset of the stream. Setting this bit indicates that the frame marks the end of the stream. Thus, this is the last message before the client sends its HTTP GET which means that the \textit{connect} duration for 0-RTT accurately reflects the time it takes for the client to send its GET request. As a result, the H3 0-RTT \textit{connect} time is a valid metric to look at while measuring how long it takes until the first request is sent.

The plot shows that there is a difference between H3 0-RTT and 1-RTT of much less than one round-trip. The median for the \textit{connect} duration of H3 0-RTT is 1.17 round-trips, which increases to 1.40 round-trips for 1-RTT (for comparison, DoQ has a median of 1.43 round-trips). However there is also a distinct step pattern visible in the distribution. While the values provided are normalized by the round-trip times for the access technologies, these steps are in fact caused by the difference between access technologies, meaning that the access technologies scale differently.

\begin{figure}[!t]
    \centering
    \begin{subfigure}[t]{1\linewidth}
        \centering
        \includegraphics[width=0.95\linewidth]{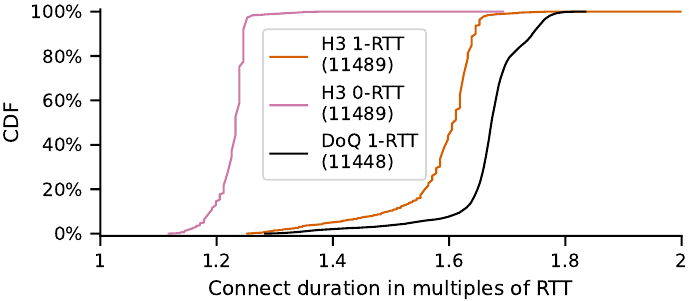}
        \caption{Fiber scenario}
        \label{fig:Image12a}
    \end{subfigure}%
    \vspace{1em}
    \begin{subfigure}[t]{1\linewidth}
        \centering
        \includegraphics[width=0.95\linewidth]{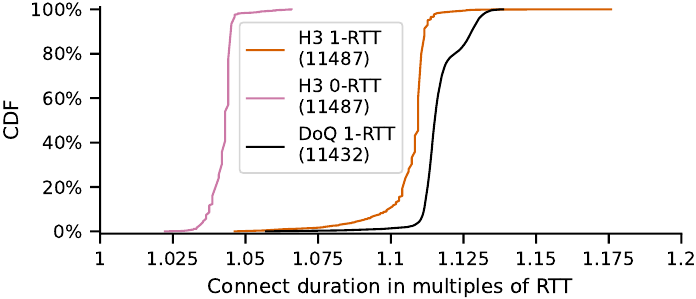}
        \caption{4G Scenario}
        \label{fig:Image12b}
    \end{subfigure}
    \caption{\small \sl CDF of the QUIC handshake connect duration H3 for 1-RTT and 0-RTT, as well as DoQ 1-RTT. For fiber, the difference between HTTP 0-RTT and 1-RTT is large because the RTT is relatively low and thus the processing delay has a higher share. For 4G, the difference between 0-RTT and 1-RTT is small compared to other access technologies because the processing delay is small in proportion to the RTT.}
\end{figure}

Figs. \ref{fig:Image12a} and \ref{fig:Image12b} reflect how the access technologies scale for fiber and 4G scenario respectively. It is observed from Fig. \ref{fig:Image12a} that the distributions for \textit{connect} times have a long tail in the high percentiles. 1-RTT shows a relatively large left tail from the minimum (i.e., 0th percentile, 1.25 round-trips) to around the 20th percentile (1.56 round-trips). We already know, the minimum for 0-RTT is 1.12 round-trips and the P20 value is 1.21 round-trips. As all data points are scaled by the same factor for a particular access technology, it means that the actual data itself for 0-RTT has less variation compared to 1-RTT. The median number of round-trips for 0-RTT is 1.23, which increases to 1.61 round-trips for 1-RTT (difference of 0.38 round-trips).

Comparing this observation to the difference in round-trips for 4G in Fig. \ref{fig:Image12b}, we observe that the median for 1-RTT increases to 1.12 from 1.06 round-trips as for 0-RTT. The plot also shows how the different steps in Fig. \ref{fig:Image11} correspond to different access technologies despite normalizing by delay. Looking at the 0-RTT distribution, the step from P0 to P20 corresponds to the data shown in Fig. \ref{fig:Image12b}. The step from P20 to P40 corresponds to 4G medium, the one from P40 to P60 is for cable, P60 to P80 is for fiber and lastly, P80 to P100 is for DSL. In addition to this, Fig. \ref{fig:Image12b} also shows that 4G handshake time scales better with RTT while having less variation, thereby covering a smaller range of values. The minimum and maximum values for 0-RTT are 1.02 and 1.07 round-trips respectively.  

\begin{summary}
\takeaway{
The overhead of client and/or server-side processing delay is relatively large for measurement setups where a low RTT access technology is emulated.
While, in absolute terms, the processing delay is the same for access technologies with high RTTs, it weighs in much less relatively, resulting in the observed differences between H3 0-RTT and 1-RTT to be small in that case.
However, 0-RTT still shows \textit{connect} times.
}
\end{summary}

\subsection{On DNS Overheads}
\label{sec:3.2}

To evaluate the overhead of DoQ and DoH in comparison to unencrypted DoUDP, we analyze the scaling factor for all the measured DNS protocols in terms of lookup times/exchange times (i.e. handshake times + query times). The data points are normalized by the scenario’s delay where the expected values are: DoUDP does not require any connection setup round-trips, and we do not find any timeouts in our measurements; hence, the complete DNS exchange should take one round-trip in total. For DoQ, we assume QUIC \texttt{Address Validation Using Retry Packets} is disabled, as existing literature ~\cite{doqWebPerf,doqpam2022,endres.2022.quic.performance} confirms that the Address is already validated by receiving a 1-RTT packet; hence, the DoQ handshake takes one round-trip. For DoQ, the handshake is without address validation which means it takes one round-trip. By adding the DNS query on top of that, DNS resolution then takes two round-trips in total. DoH is run with TLS 1.3 and thus the handshake takes two round-trips; adding the query time results in a total of three round-trips.

\begin{figure}[!t]
    \centering
    \includegraphics[scale=0.72]{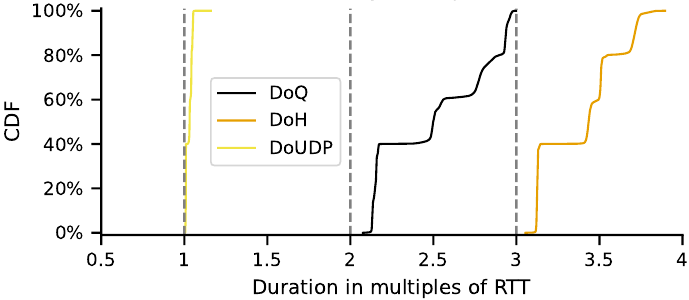}
    \caption{\small \sl CDF of DNS exchange duration in multiples of round trip times for all scenarios. Only DoUDP scales with the number of expected round-trips.}
    \label{fig:Image8}
\end{figure}

Fig. \ref{fig:Image8} shows the normalized lookups for all the three DNS protocols. It is observed from the plot that there are steps in the distribution for DoQ and DoH but not for DoUDP. The median for DoUDP is 1.03 round-trips whereas the maximum is 1.16 round-trips. For DoQ, the median is 2.50 round-trips, the minimum is 2.07 round-trips and the maximum is 3.00 round-trips. For DoH, we see this increases by almost exactly one round-trip where the median is 3.43 round-trips having a minimum of 3.05 round-trips and a maximum of 3.89 round-trips. This means that while both DoQ and DoH do not appear to exhibit the expected number of round-trips for the whole DNS lookup, the difference between them is roughly one round-trip. The five steps in 20 percentile intervals are visible for DoQ as well as DoH and represent the different access technology scenarios. Since DoUDP scales with delay as per expectation, the overhead is likely not caused by any socket setup or network stack delay.

\begin{figure}[!t]
    \centering
    \begin{subfigure}[t]{0.5\textwidth}
        \centering
        \includegraphics[width=1\linewidth]{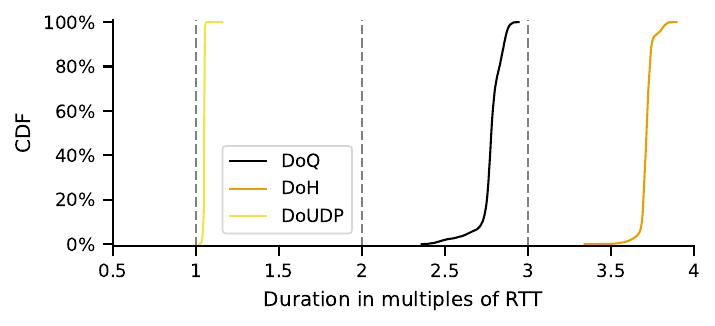}
        \caption{Fiber Scenario}
        \label{fig:Image9a}
    \end{subfigure}%
    \vspace{1em}
    \begin{subfigure}[t]{0.5\textwidth}
        \centering
        \includegraphics[width=1\linewidth]{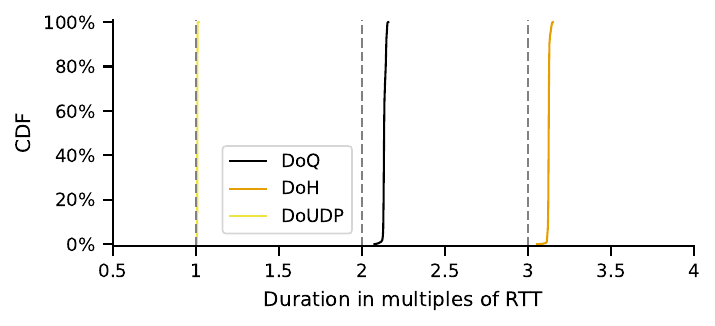}
        \caption{4G Scenario}
        \label{fig:Image9b}
    \end{subfigure}
    \caption{\small \sl CDF of DNS exchange duration in multiples of RTT. Only DoUDP scales  with the number of expected round-trips. The difference between DoQ and DoH is also one round-trip.}
\end{figure}

To confirm the above claim, Figs. \ref{fig:Image9a} and \ref{fig:Image9b} show the CDF of DNS exchange duration for the fiber and 4G setups respectively. The left tail for lower percentiles visible in the fiber plot for DoQ are also visible for DoH. The minimum (i.e., best case) for DoQ is 2.36 round-trips whereas for DoH it is 3.34 round-trips. The median, however, increases to 2.78 and 3.71 round-trips for DoQ and DoH respectively. Compared to 4G, the minimum for DoQ is 2.08 round-trips with a median of 2.13. For DoH, this increases by almost exactly one round-trip to 3.05 and 3.12 round-trips. This shows that the range of values for 4G is much smaller, meaning there is less variation in the data and there is no long tail as well. Analysing other access technology scenarios, the left tail appears to be the largest for fiber whereas it gets smaller when looking at scenarios with higher delay.

Finally, there exists one access technology where the difference between DoQ and DoH is not equivalent to one round-trip. Namely, in the case of DSL, the median of DoQ is 2.94 round-trips, while for DoH it is 3.51 round trips. This means that in this case, DoQ seems to have increased delay, despite the fact that Bandwidth-Delay Product (BDP) should be high enough. This increase is caused by higher than normal query duration. Note that the median DoQ query duration for DSL is 1.37 round-trips (min 1.35, max 1.42). For other access technologies the median is between 1 to 1.05 round-trips with no noticeable outliers for minimum or maximum values.

Digging deeper into this aspect, the measurement also contains data for the RTT of TCP (i.e., client sends a SYN and server responds with a SYN-ACK). The TCP round-trip times are analyzed to inspect whether the reason for the unusual scaling of DoH is rooted in something related to the TCP handshake or the TCP network stack itself. Since DoQ is run over UDP, the DoUDP can be used as the UDP socket setup time. The insights from above then indicate that at least for DoQ, the increased delay is not caused by anything related to the UDP stack and is likely caused by the QUIC stack.

\begin{figure}[!t]
    \centering
    \begin{subfigure}[t]{0.5\textwidth}
        \centering
        \includegraphics[width=0.95\linewidth]{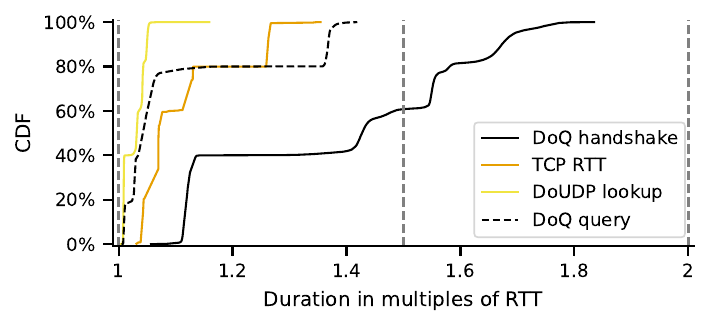}
        \caption{All Scenarios}
        \label{fig:Image10a}
    \end{subfigure}%
    \vspace{1em}
    \begin{subfigure}[t]{0.5\textwidth}
        \centering
        \includegraphics[width=0.95\linewidth]{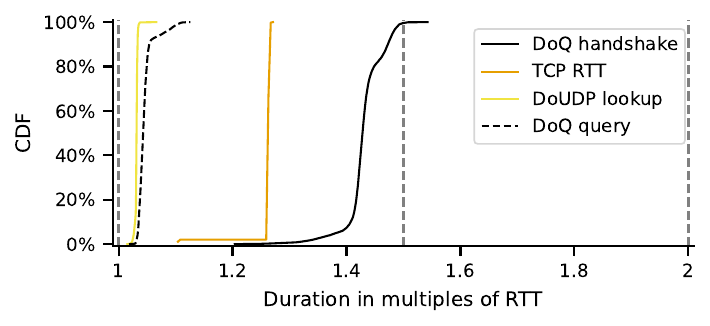}
        \caption{Cable Scenario}
        \label{fig:Image10b}
    \end{subfigure}
    \caption{\small \sl CDF of TCP RTT, DoUDP lookups, DoQ queries, and DoQ handshakes, for all Scenarios. In theory, all these metrics (except for DoQ handshake durations) should take one round-trip.}
\end{figure}

Fig. \ref{fig:Image10a} shows the TCP RTT, DoUDP lookup times, DoQ handshake times and DoQ query times. It is observed that for most of the data points, the scaling of DoUDP (median 1.03 RTTs), TCP RTT (1.07 RTTs) and DoQ query times (median 1.04 RTTs) are as expected. Explicitly, for DoQ query times, the increase for DSL is visible from P80 to P100.

There is also a noticeable increase in round-trips for this percentile range of TCP RTT. These data points belong to samples from the cable scenario, depicted in Fig. \ref{fig:Image10b}. Here TCP RTT performs worse compared to both DoUDP lookups and DoQ query times across all percentiles. It is to be noted that the minimum value for TCP RTT is 1.10 round-trips, the median is 1.26 and the maximum is 1.27. On the contrary, DoUDP is at most 1.06 round-trips whereas DoQ queries are at most 1.13 round-trips.

\begin{summary}

\takeaway{DNS over QUIC shows expected improvements over DoH due its handshake requiring less RTTs, resulting in the DNS exchange duration of DoQ being roughly one round-trip faster in comparison to DoH for all scenarios except DSL.
Moreover, lower RTT access technologies exhibit longer left tails, which eventually get smaller with increasing delay.}

\end{summary}


\subsection{On Interactions of H3 Across Different DNS Protocols}
\label{sec:3.3}

We perform experiments for three DNS protocols DoQ, DoH, and DoUDP, where DoH and DoUDP represent the encrypted and unencrypted DNS protocols commonly used in current web browsers. Each DNS protocol is combined with both H3 $0-$RTT and H3 $1-$RTT web performance measurements. A common web browsing scenario is defined as using DoUDP with H3 which is a realistic setup that likely provides the best performance with the caveat of DNS being unencrypted. DoQ with H3 $0-$RTT is referred to as QUIC \emph{connection coalescing} as it represents the emulated optimized QUIC setup. Correspondingly, DoQ with H3 $1-$RTT is referred to as DoQ whereas DoH + H3 $1-$RTT is referred to as DoH. There are also permutations of DoUDP and DoH in combination with H3 $0-$RTT which are not investigated in this paper. 

As DNS resolution is decoupled from the web browser, the DNS lookup time is added to the PLT web performance metric for H3 web performance measurement. Recall that one of our goals is to analyze how an optimized QUIC setup could perform. This is approximated by calculating the PLT for the setup where DoQ is used for DNS resolution and consequently \textit{Chromium} is used to connect to the H3 server using a QUIC $0-$RTT handshake. Such a coalesced QUIC connection would take one round-trip for the initial QUIC connection (without address validation), another round-trip for the DNS query and a third round-trip for the H3 \texttt{SETTINGS} exchange. After that the actual H3 \texttt{GET} request and corresponding response takes place. Importantly, the \texttt{SETTINGS} exchange adds a round-trip because it is not implicitly done with the initial QUIC handshake or the DNS exchange. This results in three round-trips until the client sends its GET request, which is the same number of round-trips as the non QUIC coalescing scenario with DoQ and normal H3. This means that only the processing delay for the client and the server where they know the \texttt{SETTINGS} parameters beforehand and the server not having to send its certificate twice are subtracted from the overall web performance of normal H3 with DoQ.

The first set of experiment provides an overview of the median PLT increase for all the considered access technologies and web pages. The relative increase over the DoUDP + H3 $1-$RTT baseline is calculated for three protocol combinations: QUIC connection coalescing, DoQ and DoH. The relative increase is calculated using median values for both the protocol combinations (i.e., baseline and the comparator). The measurement is performed for a specific access technology and web page combination where the web pages are ordered by complexity horizontally. Note, the \textit{example page} is a single HTML document whereas the \textit{wikipedia  page} includes Javascript in the HTML document to build the web page by fetching a single Javascript resource. On the contrary, the \textit{instagram page} requires parsing and execution of seven Javascript resources (including React.js), two style sheets and finally produces a cookie popup banner while loading. The access technology scenarios are sorted by their delay vertically.

\begin{figure*}[!tb]
    \centering  \includegraphics[width=0.90\linewidth]{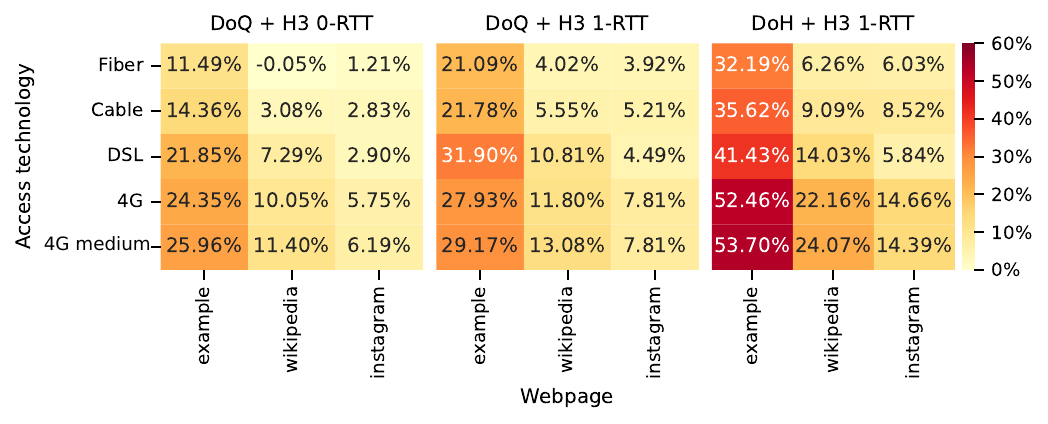}
    \caption{\small \sl Heat map of relative PLT increase over DoUDP baseline for QUIC connection coalescing (i.e. DoQ + H3 $0-$RTT), DoQ + H3 $1-$RTT, and DoH + H3 $1-$RTT. Cost of encryption is substantially reduced when encrypted connections are coalesced using DoQ + H3 $0-$RTT.}
    \label{fig:Image1}
    \vspace{-0.5em}
\end{figure*}

We observe from Fig. \ref{fig:Image1} that DoH setup has the highest relative increase across all web pages and access technologies. For the \textit{example page} over 4G medium, it also has the overall worst case relative PLT increase of 53.7\%. Additionally, for all the three protocol setups, the highest relative increase is observed for the \textit{example page}. For almost all cases the relative increase for the \textit{wikipedia  page} is comparatively greater than that of the \textit{instagram page}. This follows from the web page complexity as the \textit{instagram page} is more resource-full and render time intensive than the \textit{wikipedia  page}. 
Lastly, we observe that for a lot of the web page columns, the performance of the access technologies degrade in an order of the respective RTT (delay). However, there are quite a few exceptions to this. For example, the relative increase for the DoQ setup over the baseline for the \textit{example page} is highest in case of the DSL scenario as opposed to the 4G or the 4G medium one. On the other hand, loading the \textit{instagram page} over DSL using the DoH setup (5.84\%) observes lower relative increase than that of fiber (6.03\%).

\begin{figure*}[!htb]
    \centering  \includegraphics[width=0.90\linewidth]{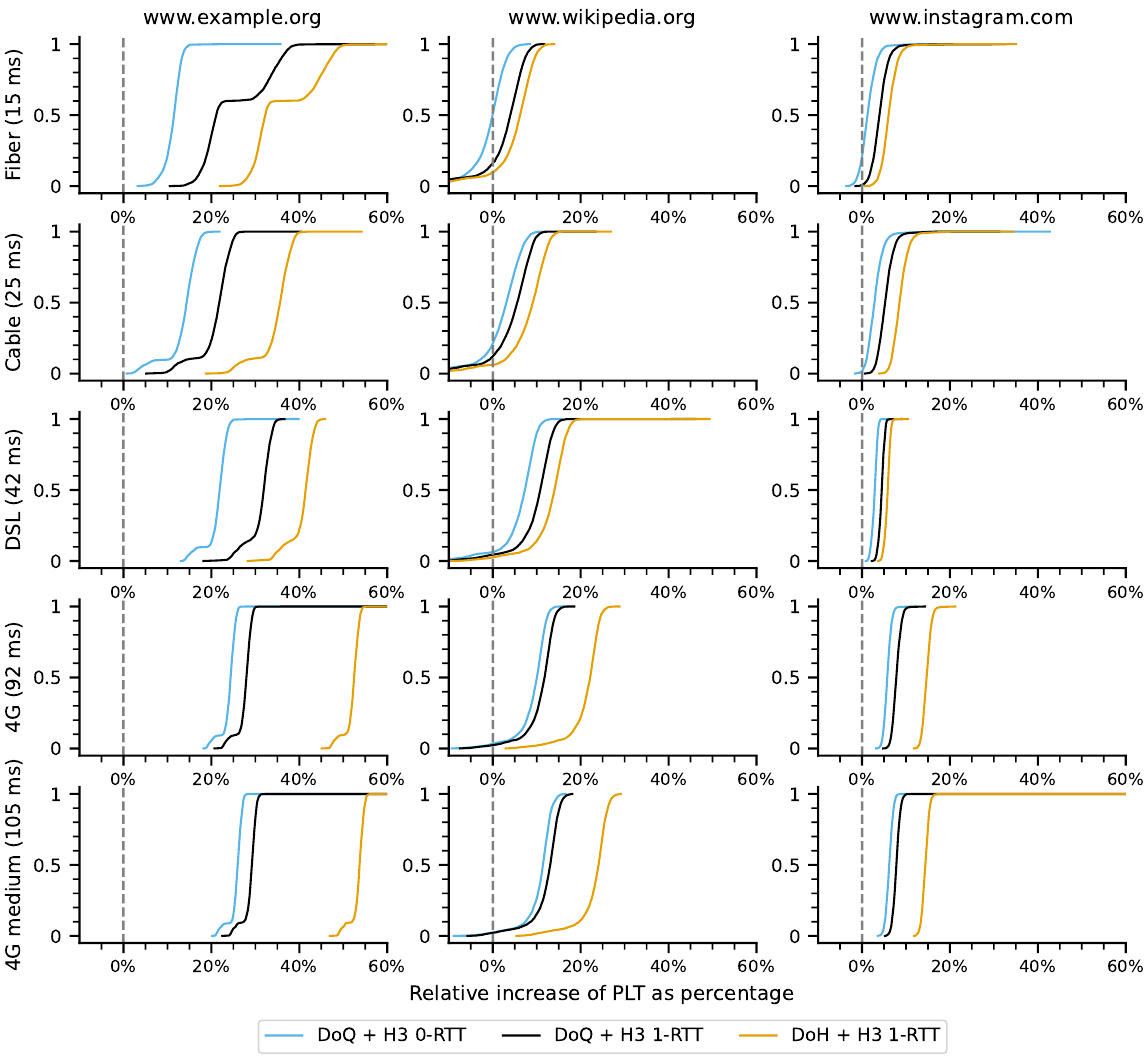}
    \caption{\small \sl Grid of CDFs showing the relative increase of QUIC connection coalescing (i.e. DoQ + H3 $0-$RTT), DoQ + H3 $1-$RTT, and DoH + H3 $1-$RTT over the horizontal DoUDP baseline for the five access technologies and the three web pages. Relative changes between the protocol combinations are affected by both of these dimensions.}
    \label{fig:Image2}
\end{figure*}

In the second set of experiment, we show the relative PLT increase in more detail. The distribution of the relative increase of all the PLTs (i.e., not just the median) over the median of DoUDP baseline are shown in Fig. \ref{fig:Image2}. Note that in theory, the relative increase can be calculated using the value of the baseline for the same measurement run, since all protocol combinations are measured in every single run. However, the advantage of using the median is that the distribution of the data points relative to each other (data point represents frequency/probability) stays the same in comparison to the distribution of the absolute PLT values.

We observe that for the fiber scenario, measuring the \textit{example page} over H3 $1-$RTT produces a distribution where there are two steps to the CDF along with two distinct PLT values that occur more frequently as opposed to a normal distribution centered around one value. This happens at the $60^{th}$ percentile, i.e., 60\% of the data points are likely centered around one PLT value and the remaining 40\% around another, higher one. To dig deeper, we investigate the other web performance metrics. It is observed from the data that this split in values is first visible for the \textit{domInteractive} metric. Before that, \textit{responseEnd} doesn't have split values. This means that the root cause behind such distinct central values is not related to fetching the web page, instead they are a result of building the Document Object Model (DOM). Additionally, this happens when \textit{gzip} is disabled and not from decoding the HTML document.

Another observation specific to the \textit{example page} is that for all access technologies excluding fiber, there is a short left tail in the distribution upto the $10^{th}$ percentile. For example, in case of cable the P10 relative increase for DoQ scenario is 14.5\%, while the P20 value is 19.6\% and the corresponding median is 21.8\%. These tails are a result of both the handshake time having left tails, as shown above along with the time it takes to fetch additional resources plus the rendering time. For example, the distributions of the time between \textit{responseEnd} and \textit{loadEventStart} has similar short left tails. For the \textit{wikipedia  page} there is a longer left tail compared to the \textit{example page} across all access technologies, however for the \textit{instagram page}, there is no left tail visible at all.

Overall, Fig. \ref{fig:Image2} demonstrates that both dimensions (i.e. web page and access technology) have an effect on the relative increase over the DoUDP baseline as well as the difference between the protocol setups. Specifically for the simplest web page, i.e. the \textit{example page}, the differences in percentage points between the protocol combinations are the largest, and for the \textit{instagram page}, the differences between them are significantly reduced. This apparently happens as the time spent by the browser in parsing the HTML documents, building the DOM and executing Javascript increases, henceforth the DNS and H3 connection setup times have less influence on the total PLT. With increasing complexity of the web page, the potential time saving (in relation to the time it takes to load a page) from changing the underlying protocols used for DNS and H3 significantly decreases.

The difference between DoQ and DoH scales with the round-trip time (except for the DSL measurement, see §~\ref{sec:3.2}). However, the difference between H3 $0-$RTT and $1-$RTT does not, as can be seen in Fig. \ref{fig:Image2} as well. For instance, observing the fiber scenario with the lowest round-trip time for the \textit{wikipedia  page}, the difference in medians between the QUIC connection coalescing setup and DoQ is 4.0 percent points. On the other hand, the difference between the medians of DoQ and DoH is 2.3 percent points. However, with increasing round-trip times (i.e., CDFs below fiber in the same column), the percentage point difference between DoQ and DoH increases. For example, in case of 4G, it increases to 10.4 percent points, while the difference in medians between DoQ and QUIC connection coalescing decreases to 1.8 percent points. The same effect is visible in the distributions for the \textit{instagram page} where fiber $0-$RTT (at the median) scenario saves 2.7 percent points while transitioning from DoH to DoQ saves 2.1 percent points. For 4G, these values are 1.6 percent points and 6.6 percent points respectively. Since all data points within a CDF are scaled by the same median value, this observation also holds for the absolute PLTs.

Overall, these observations mean that with increasing delay between the client and server, the potential time savings (relative to the PLT) of $0-$RTT decreases, while the savings for using DoQ instead of DoH increases as time spent by the browser in rendering is less affected by delay. However, it is still slightly affected by delay because of resources that need to be fetched after the base HTML document is retrieved.

\begin{summary}

\takeaway{
Using H3 $1-$RTT, page load times with DoH can get inflated by $>$30\% over fixed-line and by $>$50\% over mobile compared to unencrypted DoUDP. However, cost of encryption is substantially reduced when encrypted connections are coalesced using DoQ + H3 $0-$RTT, thereby reducing the page load times by 1/3 over fixed-line and 1/2 over mobile compared to the existing setup. Overall, our findings show that QUIC connection coalescing is the best option for encrypted communication on the Internet.
}

\end{summary}

\subsection{On a Deep Dive into Website Categories \label{sec:website_categories}}

The effects of both the dimensions (i.e. access technology and website complexity) on web performance and its scaling with respect to the underlying protocols is studied here in terms of absolute PLT values. Owing to the simplicity of the HTML page (\textit{example page}), the observed PLT distributions in the previous section are justified, hence further discussion about it is omitted. For the remaining two categories of websites, the optimal scenarios for fixed broadband (fiber) and cellular connectivity (4G with good reception) are analyzed.

\noindent \textbf{HTML Page with Javascript:} The \textit{Wikipedia  page} consists of an index HTML document (18,252 bytes), two \textit{png} logos (15,857 and 2,039 Bytes), an \textit{svg} sprite (17,229 Bytes) and a Javascript resource (614 Bytes, no compression). These byte values are not an exact match with the original \href{https://www.wikipedia.org/}{Wikipedia Page}.

\begin{figure}[!t]
    \centering
    \begin{subfigure}[t]{0.5\textwidth}
        \centering
        \includegraphics[width=0.98\linewidth]{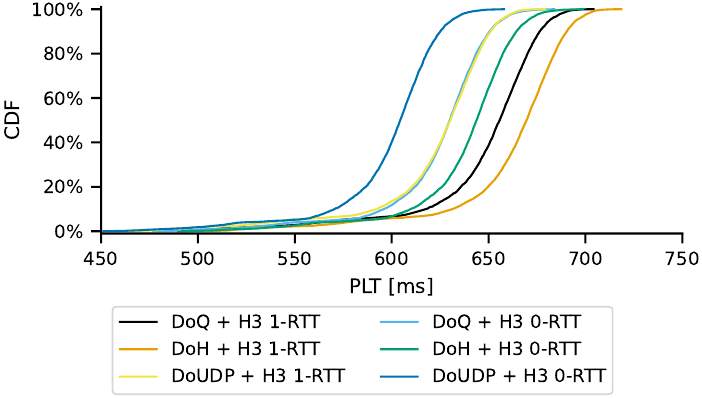}
        \caption{Fiber Scenario}
        \label{fig:Image3}
    \end{subfigure}%
    \vspace{1em}
    \begin{subfigure}[t]{0.5\textwidth}
        \centering
        \includegraphics[width=0.98\linewidth]{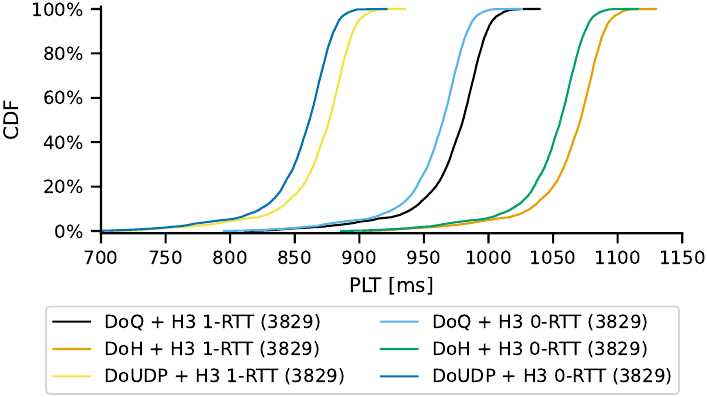}
        \caption{4G Scenario}
        \label{fig:Image4}
    \end{subfigure}
    \caption{\small \sl CDF of the PLTs for all protocol combinations for the Wikipedia  page.}
\end{figure}

Looking at a common web browsing scenario, the median is fixed to 630.4 ms which acts as the baseline for the measurement. This is also where the simulated QUIC connection coalescing setup (median 630.0 ms) matches perfectly with the baseline. Fig. \ref{fig:Image3} shows the PLTs for the fiber scenario (RTT 14.8 ms). As all the protocol combinations have a long left tail, only the median values are discussed here. It appears to be equal to the baseline across all percentiles, indicating that this is not just an artifact of the data: the baseline indeed has a P25 (Q1) value of 615.5 ms and a P75 (Q3) value of 642.1 ms, resulting in an interquartile range (IQR) of 26.6 ms. For QUIC connection coalescing, Q1 is observed to be 616.4 ms and Q3 is 641.3 ms with an IQR of 24.9 ms. The DoQ with HTTP/3 1-RTT setup has a median of 655.7 ms, which increases to 669.8 ms on changing the DNS protocol to DoH. This results in a difference of 14.1 ms between the two setups, which is almost exactly the round-trip time. This is lower than the benefit of QUIC connection coalescing over DoQ (median difference of 26 ms), which also means that the optimized QUIC setup saves more than a single round-trip time.


Figure \ref{fig:Image4} depicts the results for the same website under the 4G scenario. In this case, the DNS protocol being used has some influence on the PLT, unlike, the scenario where HTTP/3 0-RTT is used which produces lesser benefits due to the exchange of settings values. The observed median difference between 0-RTT and 1-RTT is on average 15.4 ms, which is lower than the one seen for 0-RTT fiber scenario (median difference of 26 ms). This means that benefit obtained from QUIC connection coalescing (based on HTTP/3 0-RTT) is somewhat dependant on the round-trip delay, however, the major component is still lower processing delay. The median for the QUIC connection coalescing setup is 965.0 ms while that of the baseline setup is 876.9 ms, thus, bearing a difference of 88.1 ms. Correspondingly, the DoQ setup has a median of 980.4 ms, thereby, having a difference of 103.5 ms to the baseline. Finally, the difference is median values between the DoQ and DoH setups is 90.8 ms which is highly reflected in the round-trip time (91.9 ms) for the setup as well. Overall, this means that while the QUIC connection coalescing setup does perform fairly well (excluding the baseline), maximum benefit can be gained from using DoQ over DoH.

\noindent \textbf{HTML Page with Javascript and CSS:} A major part of rendering the \textit{Instagram page} is related to building the user interface from seven Javascript resources, two style sheets, a cookie popup banner and a login form. The various images (such as the Instagram, App Store and Play store logos) embedded in the \textit{Instagram page}, are triggered by scripts loaded after the index HTML document is fetched. The website also attempts to load app screenshots, but they are never rendered in the current setup due to the viewport size being too small. However, an image of a smartphone that acts as a border around these screenshots is mirrored correctly but isn't fetched. Due to these reasons, the measured PLTs are different from that of the real website. 

%

\begin{figure}[!t]
    \centering
    \begin{subfigure}[t]{0.5\textwidth}
        \centering
        \includegraphics[width=0.95\linewidth]{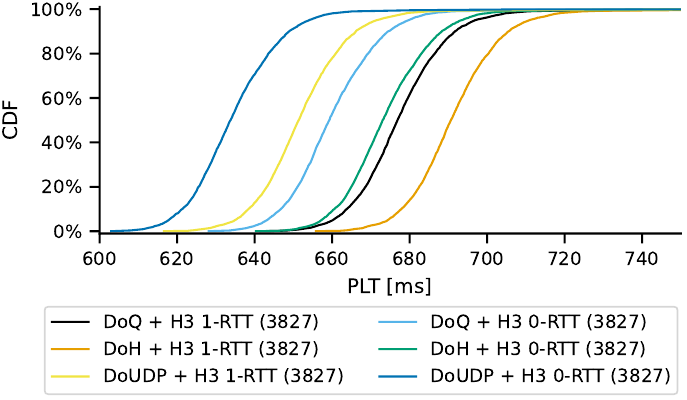}
        \caption{Fiber Scenario}
        \label{fig:Image5}
    \end{subfigure}%
    \vspace{1em}
    \begin{subfigure}[t]{0.5\textwidth}
        \centering
        \includegraphics[width=0.95\linewidth]{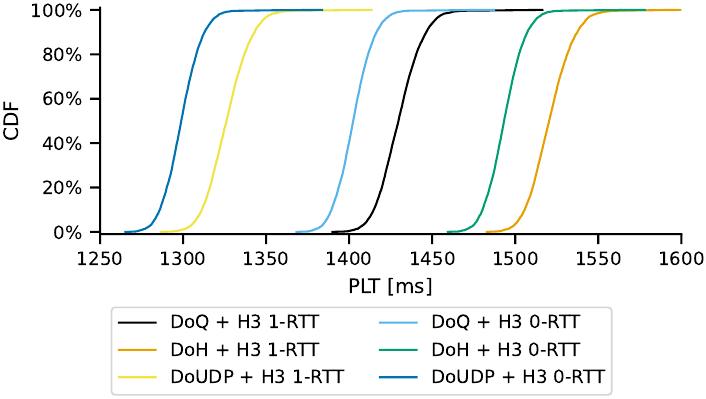}
        \caption{4G Scenario}
        \label{fig:Image6}
    \end{subfigure}
    \caption{\small \sl CDF of the PLTs for all protocol combinations for the Instagram page.}
\end{figure}

Figure \ref{fig:Image5} depicts the PLTs for the fiber scenario. It is observed that PLTs of the \textit{Instagram page} is higher compared to the \textit{Wikipedia  page} due to its greater complexity. We also observe that difference between 0-RTT and 1-RTT is relatively closer to the differences between the DNS protocols. Here, 0-RTT saves on average 17.6 ms which is slightly more than one round-trip time (14.8 ms). However, \textit{Instagram page} achieves lower savings compared to that of the \textit{Wikipedia  page} (26 ms). This implies that benefits of QUIC connection coalescing also depend on the website’s complexity. The median for the baseline is 651.4 ms while the median for the QUIC connection coalescing setup is 659.3 ms thus bearing a difference of 7.9 ms. Similarly, by comparing the DoQ setup to the baseline, we observe a difference of 25.5 ms with the median at 676.9. The difference between DoQ and DoH (690.7 ms median) is 13.8 ms which is 1 ms lower than the applied delay. Similar to the \textit{Wikipedia  page}, the potential benefit of using a QUIC connection coalescing setup is greater than the benefit of using DoQ over DoH for encrypted DNS.


Lastly, Figure \ref{fig:Image6} shows the PLTs for the 4G scenario. Quite similar to the \textit{Wikipedia  page} the difference between medians of DNS protocols now outweighs the difference between 0-RTT and 1-RTT which implies the potential benefits of QUIC connection coalescing over DoQ. The former is 90.9 ms when going from DoH to DoQ while the latter is 27.4 ms on average, which is more than the fiber scenario. This is just the reverse of the effect seen with the \textit{Wikipedia  page} where the fiber scenario had more PLT reductions in QUIC connection coalescing than 4G. The difference between the two encrypted DNS protocols is close to the delay on the connection (91.9 ms). This effect is visible across websites and access technologies which is quite obvious as DNS timings are independent from the websites being measured. 

In conclusion, using the QUIC 0-RTT handshake does not result in a speedup of one round-trip with regard to the browser sending the initial \textit{GET} request. While using 0-RTT saves some milliseconds of processing delay in the PLTs, it is however reflected differently depending on the website and access technology. For the \textit{Wikipedia  page} under the fiber scenario, QUIC connection coalescing reduces median PLT by 26 ms compared to DoQ, which decreases to 15.4 ms under 4G. For the \textit{Instagram page} on the other hand, QUIC connection coalescing saves 17.6 ms when simulating a fiber connection, which increases to 27.4 ms for 4G. Furthermore, the difference between DoUDP and DoQ appears to be higher than one round-trip, while the difference between DoQ and DoH is very close to one round-trip.
This effectively means that the QUIC connection coalescing setup has the highest impact on connections with low delay. This is because in our measurement setup it does not scale with RTT in an obvious way. Furthermore, specific DNS protocols are less important for more complex websites (assuming all resources are served by the same host), since the browser rendering takes more time and thus DNS performance has less impact.

\begin{summary}

\takeaway{
Overall, our findings show that QUIC connection coalescing is the best option for encrypted communication on the Internet, however it is more beneficial for less complex websites. Also, the performance gains vary depending on the website and access technology combination. Lastly, QUIC connection coalescing setup has the highest
impact on connections with low delay. 
}

\end{summary}

\noindent\textbf{Summary} 

\begin{figure}[!t]
    \centering
    \includegraphics[width=\linewidth]{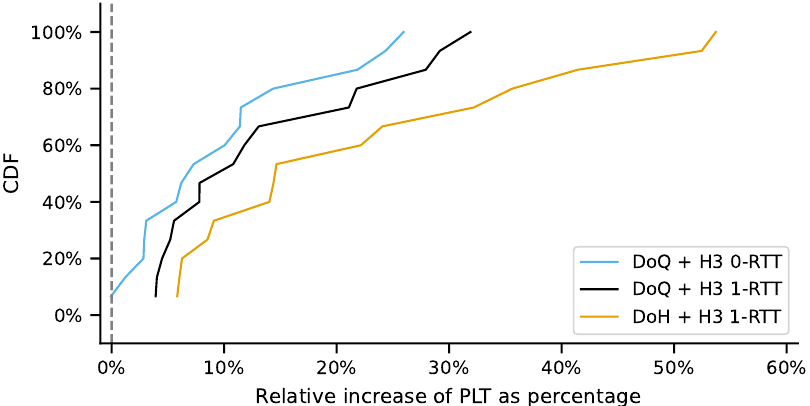}
    \caption{\small \sl CDF of the relative increase of protocol combinations over the DoUDP baseline.}
    \label{fig:Image7}
\end{figure}

To summarize, Fig. \ref{fig:Image7} shows as a CDF, the relative PLT increase (at the median) for the relevant protocol combinations to the DoUDP baseline. Each protocol combination has 15 data points in the CDF, one for each \textit{[web page, access technology]} tuple. As already explained, the baseline is a common web browsing scenario over unencrypted DNS. The QUIC connection coalescing setup can only match it for one tuple where the median relative increase is 7.3\%. For a DoQ setup, the median is slightly higher at 10.8\%. Finally the DoH setup, which is a protocol combination that is present in \textit{Chromium} right now, has an average relative increase of 14.7\%. In the worst case, QUIC connection coalescing exhibits an increase of 26.0\%, DoQ at 31.9\% and DoH at 53.7\% respectively.

The percentage point difference between DoH and DoQ in the worst case is much larger than the one between DoQ and QUIC connection coalescing. This means that for worst case scenarios, an end-user can drastically improve their performance by using DoQ. On the contrary, the end-user gains relatively less performance under a unified QUIC connection for DNS and H3. This, however, comes with the caveat that $0-$RTT does not actually save a full round-trip due to H3’s \texttt{SETTINGS} exchange. If this exchange were made earlier, e.g., by piggybacking the DNS request and response or even the initial QUIC handshake, a full round-trip could be saved, thereby making the performance closer to the baseline DoUDP + H3 1-RTT setup. However, out of the encrypted DNS protocols, QUIC connection coalescing setup is still the best option for a fast private web browsing experience. While we agree that in some use cases with browsers, e.g. static connection and same destination (hostname), the benefit of 0-RTT could be limited to the first visited website, several other realistic scenarios, e.g., end users changing the access network, e.g., moving from mobile to WiFi and vice-versa in a smartphone, closing the application (browser) or reaching a different Point-Of-Presence of a content provider, could benefit from having 0-RTT.


\section{Discussion \label{sec:discussion}}
There can be additional challenges, when QUIC is used to coalesce name resolution via DoQ and Web content delivery via H3 with 0-RTT over the same edge server. Next, we would like to comment on some aspects when it comes to performance and privacy in the current state of the Internet and for its future, with an eventual adherence to coalescence of DNS, QUIC and H3, specially.

\subsection{Connection Coalescence in the Internet}

Popular Internet content providers such as CDNs nowadays host DNS services, even their own DNS as a service, to offer content from third-party customers. From a CDNs’ perspective the clear benefit of connection coalescence is the drastic reduction in the number of parallel connections needed by a web browser to locate (DNS) and fetch (website) content arriving at their infrastructure. For instance, to read the content of a simple blog entry today, this could easily involve tens of DNS requests to resolve another tens of hostnames, serving (sub-)subresources to construct the website as part of the operation. Again, from content serving part of the network, this means a significant load per client arriving at the servers and potentially causing scalability issues. On the other hand, one can argue that in all the involved requests that much of the end user’s metadata in this operation, e.g., via the absence of an encrypted Server Name Indicator (SNI), is clearly exposed to the network (not only the CDN). With the SNI, an on-path eavesdropper could easily fingerprint the traffic coming from the end user and determine the interactions on the website. From the end user perspective coalescing the connections into a single means placing a strong level of trust in the CDN, which will have access to all data from the DNS requests to the website(s) interactions. The idea of coalescing connections is not new, it was brought from earlier HTTP versions, and most modern browsers already reuse open connections to the same website, assuming the connection identification and certificate are the same. For browsers, the clear benefit is to reduce the large number of parallel connections that can be open and connection management. Also, different forms of scheduling requests and parallelization routines can be done at the browser since there is better visibility in how content is being delivered.

From our measurements, the best option for end users today if connection coalescence was already a reality, is the clear reduction of necessary RTTs from the DNS request to the content retrieval. However, QUIC has matured over the years to show not only performance but flexibility benefits as a transport layer protocol to encrypt Internet communication. We can give a different example, which QUIC can be used as the core protocol to address concerns around privacy of users in general: Many Internet players actually need some information exposure from protocols to perform tasks such as network management. Think, for instance, of an ISP that needs some understanding about a sudden traffic influx: Is it malicious or an unexpected user activity routed to a different network segment due to an outage? With QUIC, a good portion of this traffic is invisible to the ISP that is not able to decrypt the end-to-end connection. To address such concerns, often called ``pervasive encryption”, there are solutions such as Multiplexed Application Substrate over QUIC Encryption (MASQUE), which proposes QUIC as a substrate to tunnel any type of traffic. Beyond this initial goal, it also attempts to create a collaborative approach, e.g., via a MASQUE proxy, where pieces of the end-to-end communication can be selectively exposed to certain parties, i.e., the ISP, along the end-to-end path. Beyond the demonstrated benefits of QUIC’s built-in features to carry multiple streams in a single and secure connection, it is flexible enough to also address other privacy concerns.

\subsection{Web Security, User Privacy and Trust}
While the trend of secure encrypted communication in the Internet with TLS is overall positive, the encryption of DNS requests has been heavily debated due to the privacy implications side effects imposed to end users: The DNS infrastructure has long been centralized and, by enabling encryption with approaches discussed in this paper such as DoH or DoQ and DoT, it delivers all user data in the hand of a few Internet players, i.e., hyper-giants. Although the encryption of DNS requests is unquestionably positive, it has clear consequences to performance, privacy, competition, and availability of DNS. A growing concern is the difficulty in the control and choice of the DNS recursive resolver. For instance, in a desperate attempt to lose DNS queries’ visibility, it is known that ISPs partnered with web browsers to become trusted DNS resolvers. There are a few solutions from distributing DNS queries in different ways, e.g., hash-based or randomly, to different recursive resolvers to run a proxy under control of the end user, where they can configure the “visible” parts of their DNS requests to the outside Internet.

With the deployment of connection coalescing, an increasing amount of user data will be delivered to large content-delivery hyper-giants. As such, the system can still be viewed as a trade-off between performance and privacy, depending on the  sensitivity of the user, as connection coalescing can lead to centralization of trust \cite{viet:toit:2022}. With the EU, via the Digital Services Act, working towards ensuring that hyper-giants adhere to stricter privacy and transparency obligations, the concept of Web privacy within the context of hyper-giants will continue to evolve in the coming years.
Along the lines with ISPs partnering with web browsers to become their trusted DNS recursive resolver, the same ISPs or other networks used more as transport networks, i.e., shuffling traffic, such as mobile operators, can directly collaborate with the hypergiants so that portions of the traffic stays within the ISPs. This collaborative approach is definitely more promising to improve privacy for Internet users rather than circumventing encryption or blocking traffic.

While, the presented method prevents an intruder from plain eavesdropping the browsing behaviour of the clients and/or launching man-in-the-middle attacks, whether pattern inference of encrypted packets using machine learning methods can reveal parts of browsing behavior such as with website fingerprinting (WFP) attacks \cite{sandra:ndss:2020} is an area of further exploration. 
It is technically possible to censor HTTP traffic, since the SNI option in TLS is not encrypted and still visible in H3. This could be one approach to identify and deliberately block QUIC traffic. The built-in encryption of QUIC makes it less vulnerable to other types of censorship techniques such as connection tracking, since the connection identification values change during the time of the connection. Also, even when IP addresses may change, the same QUIC connection can continue to exist.

Within this context, it would further be interesting to investigate how the performance varies in the presence of background traffic coupled with a WFP attack. Further, the presented method does not yet prevent the association of a user (via source IP address) from the requested content (via destination IP address). As such, combining the method with private relays, such as MASQUE, that leverage QUIC as the underlying protocol and is currently under standardization at the IETF, would be an interesting new direction to tighten the privacy properties of the system.

\vspace{1em}
\section{Limitations and Future Work}
\label{sec:limitations}

There are a few noticeable limitations.
First, the presented findings represent an emulated setup where the DNS name resolution had to be decoupled from the web browsing process. As a consequence, the performance metrics are computed by summing DNS time and HTTP time. Taking this factor into account, the evaluation shows a lower bound of the possible performance benefits of coalescing.
Secondly, the use case of measuring an HTML page over an emulated fiber connection shows that the page load times have two central values.
While considering all web performance metrics, we find that this split happens after the web page is already fetched while building the DOM.
Yet, we were not able to investigate the root cause of this behavior. The  measurement setup to evaluate QUIC connection coalescing using DoQ + H3 for $0-$RTT is limited to web pages having a single DNS resolution. As such, the setup itself is currently implemented with a single H3 web server that serves as a directory to replay web pages.
However, all resources being served by the same host is an uncommon scenario on the Web, since most web pages use third-party resources. Moreover,  for websites with several DNS resolutions, a scaling factor can be applied to the results presented in the paper.

We plan to further refine the introduced concept of QUIC connection coalescing in the future. For instance, \textit{Chromium} will be extended with support for DoQ in order to couple DNS resolution with web browsing, resulting in a measurement setup capable of QUIC connection coalescing.
This will also extend the methodology to web pages with more than one DNS resolution, enabling the measurement of arbitrary web pages. We also plan to extend the setup to emulate packet loss and
cross-traffic network conditions.
Finally, while we use DoH with HTTP/2 as the current de-facto standard for encrypted DNS on the web, DNS over HTTP/3 (DoH3) is expected to gain traction in the coming month.
Though not widely supported, Google has added DoH3 to their public DNS service as well as Android in July 2022~\cite{android-doh3}. Cloudflare has also added DoH3 support to their public DNS service in March 2022~\cite{cloudflare-ddr}.
Hence, we plan to extend our work with DoH3 further by blurring the boundaries between DNS resolution and Web content delivery.

\section{Conclusion}
\label{sec:conclusion}

In this paper, we evaluated the cross-layer interactions of QUIC, DNS, and H3, highlighting the benefits of using QUIC to coalesce name resolution via DNS over QUIC and Web content delivery via H3 with $0-$RTT.
With the introduced measurement setup, we performed automated measurements of DNS resolution and Web browsing while emulating network conditions based on real-world datasets for both fixed-line and mobile-access network technologies.
Our findings show that page load times using DNS over HTTPS can get inflated by $>$30\% over fixed-line and by $>$50\% over mobile when compared to unencrypted DNS over UDP, reflecting the cost of encrypted DNS.
Taking \textit{Web Privacy By Design} to the next level, we coalesced DNS over QUIC and H3 $0-$RTT connections. With reduced page load times by 1/3 over fixed-line and 1/2 over mobile compared to existing Web browsing setup, our findings highlight that QUIC connection coalescing is currently the best option for encrypted communication on the Internet.


\section*{Acknowledgment}

This work was supported by the Volkswagenstiftung Niedersächsisches Vorab (Funding No. ZN3695).

\bibliographystyle{IEEEtran}
\footnotesize{
\bibliography{References}}


\end{document}